\documentclass{article}

\usepackage[preprint, nonatbib]{neurips_2022}
\usepackage[colorlinks=true,citecolor=blue]{hyperref}
\usepackage{physics}
\usepackage[backend=biber,sorting=none,style=ieee,citestyle=numeric-comp]{biblatex}
\usepackage{comment}

\usepackage[utf8]{inputenc} 
\usepackage[T1]{fontenc}    
\usepackage{hyperref}       
\usepackage{url}            
\usepackage{booktabs}       
\usepackage{amsfonts}       
\usepackage{nicefrac}       
\usepackage{microtype}      
\usepackage{xcolor}         
\usepackage{graphicx}
\usepackage{float}
\usepackage{wrapfig}
\usepackage{amsmath,amssymb}

\title{Machine-Learning Compression for Particle Physics Discoveries}

\author{Jack H. Collins$^{1}$, Yifeng Huang$^{2}$, Simon Knapen$^{3,4}$, Benjamin Nachman$^{3,5}$, and Daniel Whiteson$^{2}$\\
{}\\
$^{1}$SLAC National Accelerator Laboratory\\
$^{2}$Department of Physics and Astronomy, University of California, Irvine\\
$^{3}$Physics Division, Lawrence Berkeley National Laboratory\\
$^{4}$Berkeley Center for Theoretical Physics, University of California, Berkeley\\
$^{5}$Berkeley Institute for Data Science, University of California, Berkeley \\
{}\\
\texttt{jackadsa@gmail.com}, \texttt{yifengh3@uci.edu}, \texttt{smknapen@lbl.gov},\\
\texttt{bpnachman@lbl.gov}, \texttt{daniel@uci.edu}
}

\begin{document}

\maketitle

\begin{abstract}
In collider-based particle and nuclear physics experiments, data are produced at such extreme rates that only a subset can be recorded for later analysis. Typically, algorithms select individual collision events for preservation and store the complete experimental response.
A relatively new alternative strategy is to additionally save a partial record for a larger subset of events, allowing for later specific analysis of a larger fraction of events.
We propose a strategy that bridges these paradigms by compressing entire events for generic offline analysis but at a lower fidelity.  An optimal-transport-based $\beta$ Variational Autoencoder (VAE) is used to automate the compression and the hyperparameter $\beta$ controls the compression fidelity.  We introduce a new approach for multi-objective learning functions by simultaneously learning a VAE appropriate for all values of $\beta$ through parameterization.  We present an example use case, a  di-muon resonance search at the Large Hadron Collider (LHC), where we show that simulated data compressed by our $\beta$-VAE has enough fidelity to distinguish distinct signal morphologies.
\end{abstract}

\section{Introduction}

The rate and size of interaction events at modern particle and nuclear physics experiments typically prohibits storage of the complete experimental dataset and require that many interaction events be discarded in real time by a {\it trigger} system.  For example, at the Large Hadron Collider (LHC), collisions occur at a rate of 40 MHz, but the ATLAS and CMS experiments recording rates are typically $\mathcal{O}(\text{kHz})$~\cite{ATLAS:2016wtr,CMS:2016ngn}.  For selected events, the complete experimental response is preserved for later analysis. When the scientific goals only require identifying events which contain rare and easy-to-identify objects, such as high energy photons, the trigger system is highly efficient.  However, this strategy leaves the vast majority of the events unexamined, including many with complex features  that are hard to quickly identify online or may not be rare.

An alternative approach to fully recording a small fraction of the events is to preserve a partial record of a larger fraction~\cite{Aaij:2016rxn,CMS:2016ltu,ATLAS:2018qto}.  This strategy has allowed access to lower-energy phenomena which occur at higher rates, but the utility of these partial data records is limited.  For example, a recent partial-event analysis targets di-muon resonances~\cite{CMS:2019buh},  only recording the four-momenta of the two muons and a small number of additional event properties for low-mass events that would otherwise be too high rate for the full-event trigger system.  This approach has the potential to make a major discovery, but the lack of a full event record could make it  challenging to \emph{diagnose} such a discovery. To distinguish between several competing hypotheses which might generate a peak in the di-muon spectrum would require recording new data with a dedicated trigger, which is both time consuming and expensive. 

We propose an approach that bridges the full and partial event paradigms automatically with machine learning.  This is accomplished by training a neural network to learn a lossy event compression with a tunable resolution parameter.  An extreme version of this approach would be to save every event at the highest resolution allowable by hardware (see e.g. Ref.~\cite{DiGuglielmo:2021ide} for autoencoders in hardware).  We present a more modest version in which we envision  full event compression which could run alongside partial event triggers to expand their utility for a larger range of offline analyses.  Our approach uses a optimal transport-based Variational Autoencoder (VAE) following Ref.~\cite{Collins:2021pld}.

In a proof-of-concept study, we compress and record a sample of simulated interactions which are similar to those analyzed in Ref~\cite{CMS:2019buh},  preserving information which would otherwise be lost. We show that this additional information can be used to effectively discriminate between two signal models which are difficult to distinguish with only the muon kinematics. The overall structure of the proposal is that first, a signal is discovered in a trigger-level analysis such as this dimuon resonance search. Subsequently, a compressed version of the hadronic event data, which has been stored alongside the muons, can be used to rule out or favor candidate signal models.

\section{Related Work}

An alternative to compressing individual events is compressing the entire dataset online~\cite{Butter:2022lkf}, which is methodologically and practically more challenging.  An alternative to saving events for offline analysis is to look for new particles automatically with online anomaly detection~\cite{Cerri:2018anq,Knapp:2020dde,Govorkova:2021hqu,Mikuni:2021nwn}.  While we build our VAE on the setup from Ref.~\cite{Collins:2021pld} using the Sinkhorn approximation~\cite{sinkhorn,sinkhorn1967} to the Earth Movers Distance, other possibilities have been explored, such as using graph neural networks~\cite{Tsan:2021brw}.  We leave a comparison of the power of different approaches to future work.

\section{$\beta$-parameterized Variational Autoencoder}

We represent each collider event $x$ as a point cloud of 3-vectors $\{p_{\textrm T}/H_{\textrm T}, \eta, \phi\}$, where $\eta$ and $\phi$ are the geometric coordinates of particles in the detector, and $p_{\textrm T}$ their transverse momenta which correspond to the weights in the point cloud. These are normalized for each event using $H_{\textrm T} = \sum_{i} p_{\textrm{T},i}$. We build an EMD-VAE~\cite{Collins:2021pld,Fraser:2021lxm, Kingma2014AutoEncodingVB} trained to minimize a reconstruction error given by an approximation to the 2-Wasserstein distance between collider events $x$ and reconstructed examples $x'$, with loss function

\begin{equation}
    L = \left<S(x,x'(z))/\beta + D_\text{KL}(q(z|x)||p(z))\right>_{p(x)}.
\end{equation}

An encoder network maps the input $x$ to a Gaussian-parameterized distribution $q(z|x)$ on 256-dimensional latent coordinates $z$. This network is built as a Deepsets/Particle Flow Network (PFN)~\cite{zaheer2017deep, Komiske:2018cqr}. A decoder $x'(z)$ maps latent codes $z$ to jets $x'$, parameterizing a posterior probability 

\begin{displaymath}
\log p(x|z) \propto S(x,x'(z))/\beta\,,
\end{displaymath}

where $S(x,x'(z))$ is a sharp
Sinkhorn~\cite{sinkhorn1967,Cuturi2013SinkhornDL,luise2018differential,patrini2020sinkhorn} approximation to the 2-Wasserstein distance between event $x$ and its decoded $x'$ with ground distance given by $M_{ij} = \Delta R^2_{ij} \equiv (\eta_i - \eta_j)^2 + (\phi_i - \phi_j)^2$, and calculated using the same algorithm and parameters as in Ref~\cite{Collins:2021pld}. This decoder network is built as a dense neural network. $D_\text{KL}(q(z|x)||p(z))$ is the KL divergence between the encoder probability $q(z|x)$ and the prior $p(z)$, which we take to be a standard Gaussian. This KL divergence can be expressed as a sum of contributions from each of the 256 latent space directions. The details of the architecture  is described in the Appendix.

The quantity $\beta$ is typically taken to be a fixed hyperparameter of the network~\cite{Higgins2017betaVAELB} which controls the balance between reconstruction fidelity and degree of compression in the latent space. In this work, we elevate $\beta$ from a fixed hyperparameter to an input~\cite{Baldi:2016fzo} of both the encoder and decoder networks\footnote{The authors are grateful to Jesse Thaler for this suggestion.}\footnote{Note added post-publication: A similar idea was pursued in \cite{bae2022multi}, which was submitted for publication concurrently with this work. The implementation in their study differs from ours by using a hypernetwork to model the $\beta$ dependence rather than embedding, which will result in differing performance characteristics that deserve further study. However, the goal is the same: to model the full rate-distortion curve with a single VAE network.}. %
Training events are each accompanied by a value for $\beta$ generated from some arbitrary distribution that has support over the region of interest. These sampled $\beta$ are provided as an additional input to both the encoder and decoder, and included in the loss calculation for each event. The encoder and decoder then become dependent on the input value for $\beta$, and can be written as $q(z|x;\beta)$ and $x'(z;\beta)$, respectively. During testing, the encoder and decoder can be evaluated on a given event with any desired value for $\beta$ as appropriate  for a particular application. In this work, we take advantage of this property purely for its usefulness in prototyping. However, it is conceivable that the ability to vary $\beta$ on the fly could prove valuable for an online application that has time-varying constraints on data transmission rates or requirements on reconstruction fidelity. Furthermore, we have found  that $\beta$-annealing is neither required nor helpful in improving our VAE performance, significantly reducing the complexity and time required for training and experiments.

\section{Numerical Results}
\label{sec:num_res}

In this section, we  illustrate how a VAE trained to compress the full event at trigger level might be used to augment a trigger-level dimuon resonance search and allow physicists to distinguish between  different potential hypothetical signal models without having to record new data using the full-event trigger system.

We train our $\beta$-parameterized VAE using simulated Standard Model events, specifically an inclusive $b$-quark jet sample ($p_{\textrm T}\gtrsim 4$ GeV/c), weighted as $\sim 1/p_{\textrm T}^{3}$. We then consider two potential signals which might be found in a trigger-level dimuon resonance search: first, a light scalar decaying to muons ($S\to 2\mu$) produced in an exotic $B$-meson decay ($B\to K S$) and secondly, a dimuon resonance produced in a hidden valley model \cite{Strassler:2006im}, from an example in Ref.~\cite{Knapen:2021eip}. The latter model has more hadronic activity than the former, as well as a small amount of invisible momentum, but has otherwise very similar muon kinematics. More details can be found in the Appendix. All signal and background events were generated with the \textsc{Pythia} event generator \cite{Sjostrand:2014zea}.

To assess how well the two models can be distinguished using only the two highest-$p_\textrm{T}$ muons, a simple binary classifier (see Appendix) was trained with only these inputs. It achieves an AUC of 0.71, indicating that the dimuon spectra are different but not convincingly so. Any attempt to distinguish the two signals based on the dimuons alone would be additionally complicated by the uncertainties on details of the showering model for the hidden valley scenario. However, the hadronic activity in the two signal classes is very distinctive, and even a relatively low-fidelity VAE reconstruction may tell them apart.

\begin{figure}[h!]
    \centering
    \includegraphics[width=0.99\textwidth]{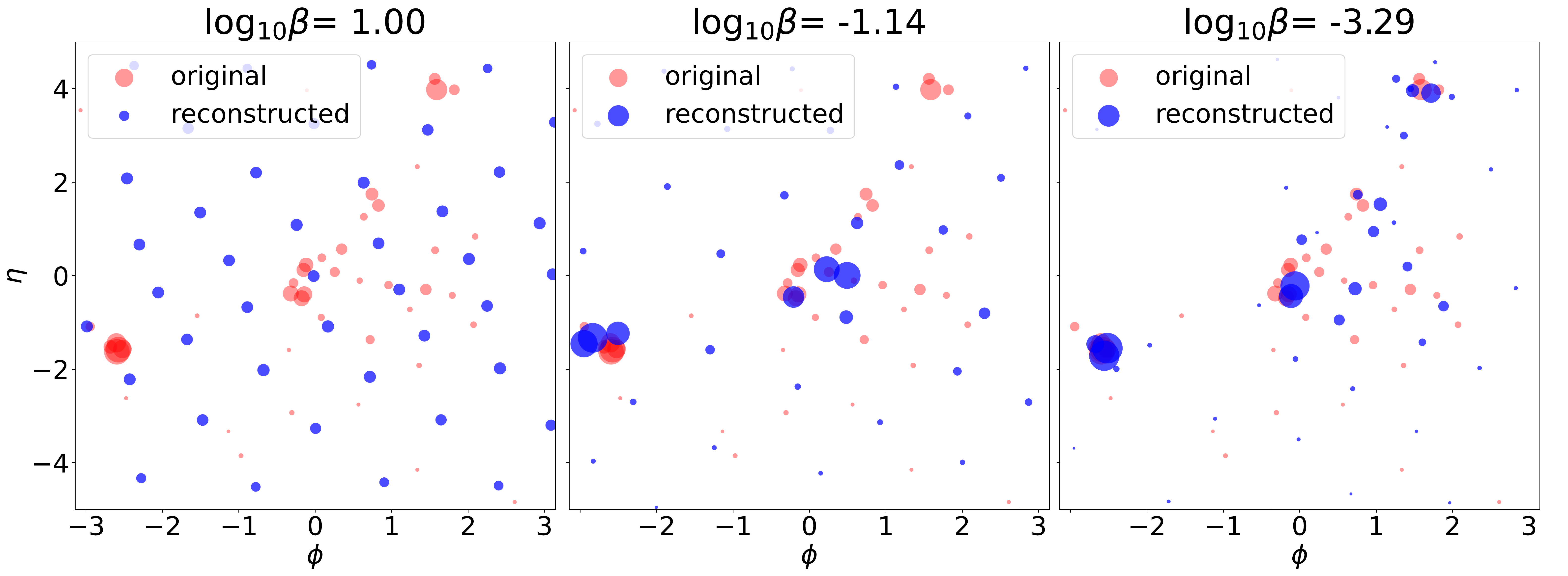}
    \caption{Reconstruction of an SM event as a function of $\log_{10} \beta$. Red points: Example test event. Blue points: MLE reconstruction of the same test event. Each point represents a particle in the event, with area proportional to the particle's $p_{\textrm T}$. The number of active latent dimensions (defined as those with $D_{\text{KL},i} > 0.1$) is 0, 9, and 48 at each of these $\beta$ values respectively, and the compression rates $D_\text{KL}$ are 0, 12, and 145 bits, respectively. }
    \label{fig:reconstructions}
\end{figure}

\begin{figure}[h!]
    \centering
    \includegraphics[width=0.45\textwidth]{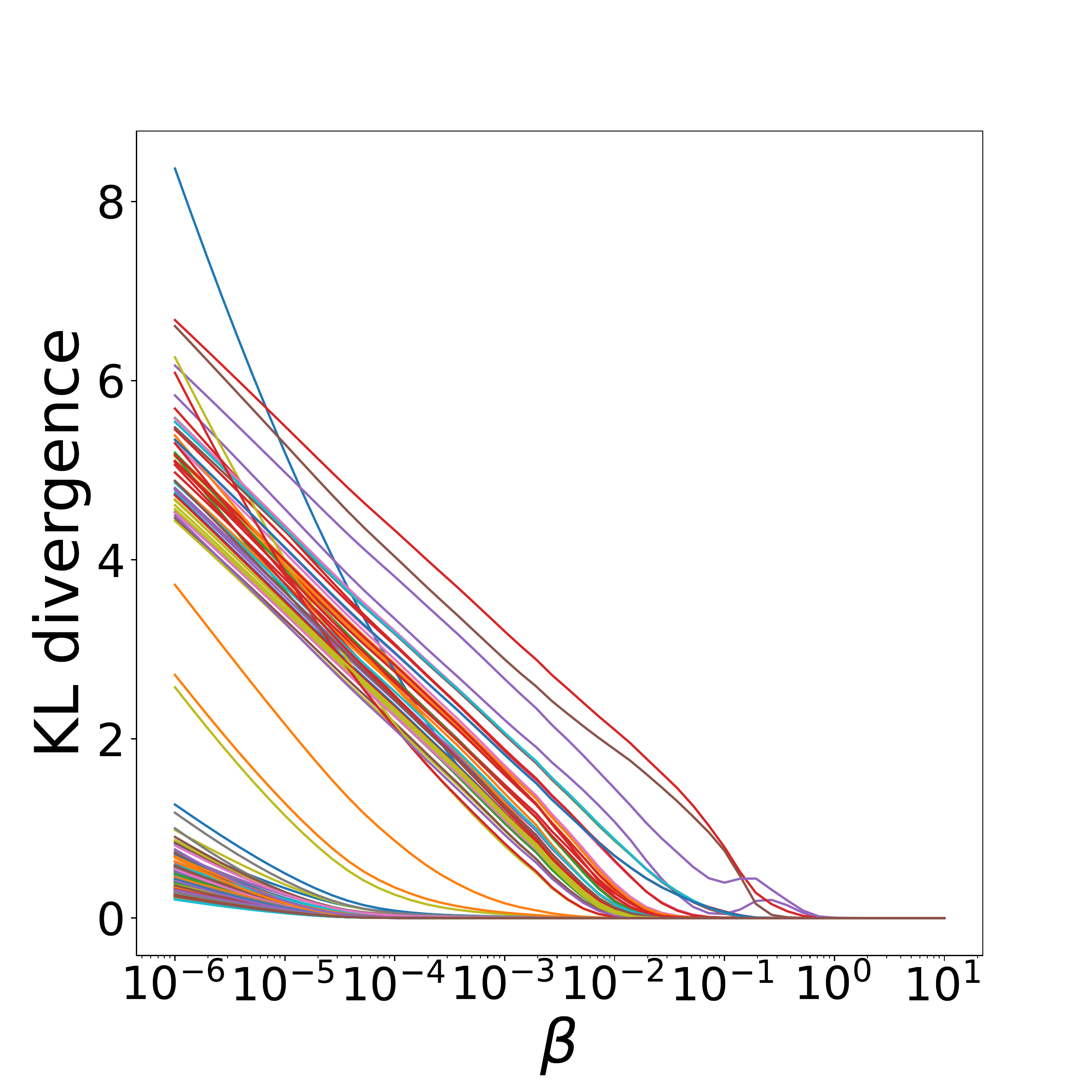}\hspace{1cm}
    \includegraphics[width=0.45\textwidth]{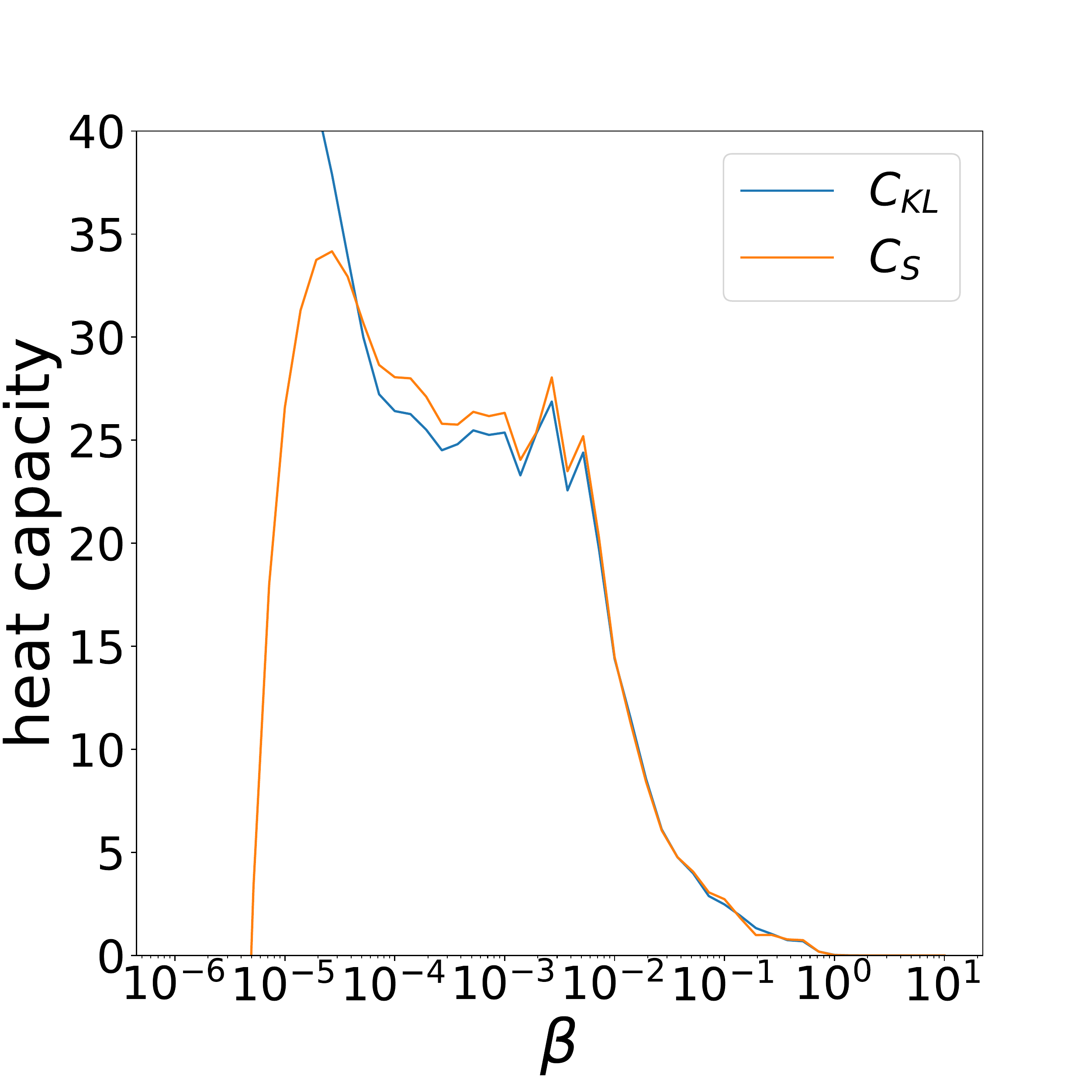}
    \caption{Left: KL divergences of the 256 individual latent space directions as a function of $\beta$, averaged over the test SM data. Right: Heat capacities on same data, defined in the text.}
    \label{fig:KLs}
\end{figure}

The structure that is learnt by the VAE as a function of $\beta$ is qualitatively illustrated in Figs.~\ref{fig:reconstructions} and~\ref{fig:KLs}. In Fig.~\ref{fig:reconstructions}, the red points represent an event from the $b\bar{b}$ test sample, and the blue its maximum likelihood estimate (MLE) reconstruction (defined using the the maximum likelihood latent code for reconstruction rather than a random sample) at various values for $\beta$. Each point represents a particle in the event, with area proportional to the momentum $p_{\textrm T}$ of the corresponding particle. We see that for large $\beta$, the VAE learns an uninformative average over all training events, but for smaller values of $\beta$ it begins to learn a more precise reconstruction.

Some properties of the information content of the learnt representations are reflected in Fig.~\ref{fig:KLs}. In the left pane, we show the individual KL divergences associated with each of the 256 latent space directions as a function of $\beta$. We see that for $\beta> 1$, all latent space directions are uninformative. Latent directions start to become sequentially informative with $D_{\text{KL}} > 0$, for $\beta < 1$. For any value of $\beta$, there is a clear hierarchy of information content in different latent directions, which can be utilized for an efficient encoding: those values which are encoded with low resolution (large variance for $q(z|x)$) can safely be represented with lower precision numbers than those having higher resolution. Note that the $\left<D_\text{KL}(q(z|x) || p(z)) \right>_{p(x)} > \mathcal{I}(x;z)$, with $\mathcal{I}$ being the mutual information, and therefore the KL divergence places an upper bound on the minimum number of bits required for the lossy compression algorithm with average distortion $\left<S(x,x')\right>$.

The overall structure of the left plot is summarized by the two heat capacities in the right plot, defined in analogy with the thermodynamic heat capacity~\cite{alemi2018therml,rezende2018taming,Collins:2021pld} by
\begin{equation}
    C_S = \frac{d\left<S(x,x')\right>}{d \beta}, ~~~~~ C_{\mathrm{KL}} = -\frac{d \left<D_\text{KL}(q(z|x)||p(z))\right>}{d \log \beta}.
\end{equation}
Because the lines in the left plot have gradient close to $-0.5$, the heat capacities are related to the effective number of degrees of the system by $\text{dim} \; \simeq 2 C$, similarly to a thermodynamic system with quadratic Hamiltonian. There is a plateau for $\beta \lesssim 10^{-2}$, indicating that the VAE is unable to learn additional informative structure at smaller scales than this. For $\beta \lesssim 10^{-4}$, the VAE begins to overfit the data and the two heat capacities diverge.

\begin{figure}[h!]
    \centering
    \includegraphics[width=0.4\textwidth]{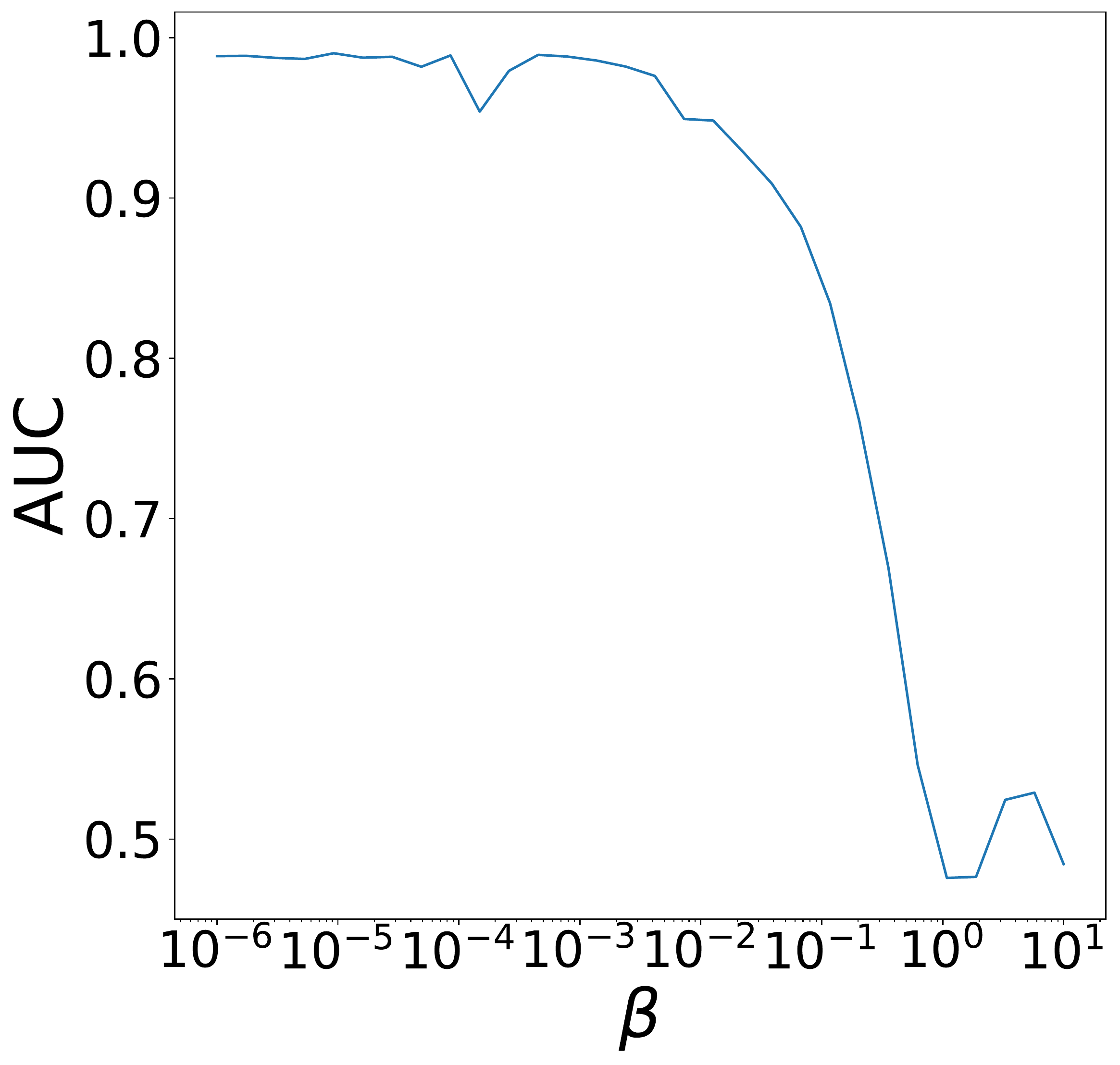}\hspace{1.7cm}
    \includegraphics[width=0.4\textwidth]{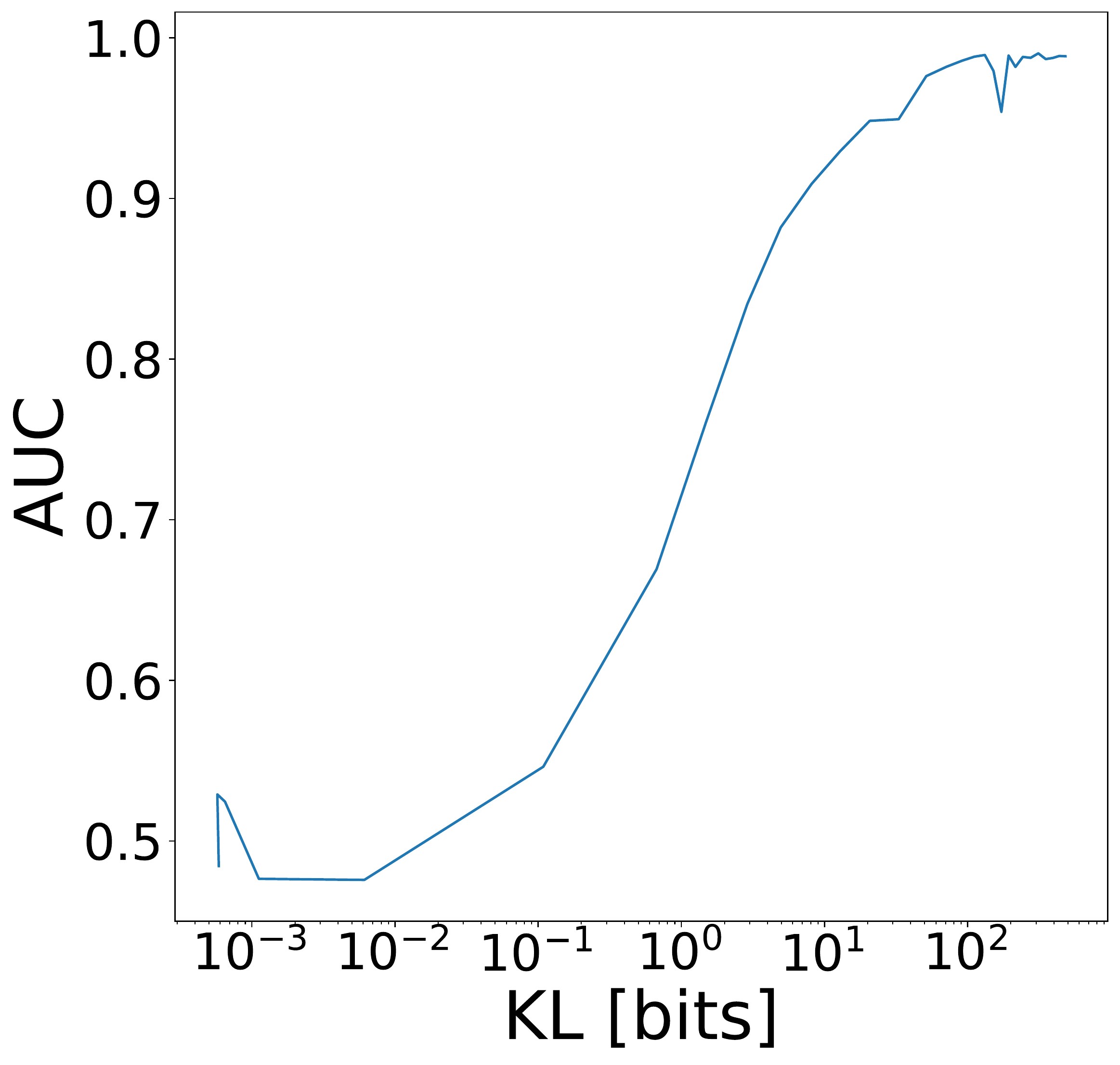}
    \caption{AUC for classification between the two signal models as a function of $\beta$ (left), and as a function of the KL divergence measured in bits (right), which reflects the compression rate.}
    \label{fig:auc}
\end{figure}

Having trained a VAE on SM data, we  explore the distinguishability of our signal models after being compressed and reconstructed.  A PFN classifier is trained on the $H_{\textrm T}$-normalized VAE reconstructions of the two signal categories. The details of this architecture and training parameters are described in the Appendix.

In Fig.~\ref{fig:auc}, we see the AUC for classification between reconstructed signal events as a function of $\beta$. For $\beta > 1$, the VAE reconstruction is uninformative for signal classification, but it contains increasingly useful information for classification for $10^{-2} < \beta < 1$, before a plateau is reached for $\beta < 10^{-2}$. This corresponds with the features in Fig.~\ref{fig:KLs} right, which shows a plateau in the effective number of degrees of freedom of the learnt representation below $10^{-2}$. These features could be combined with the dimuon resonance information (which alone provides an AUC of 0.71), and the overall $H_{\textrm T}$ of the events (which alone provides an AUC of 0.84) to further improve separation between the signals.

\section{Conclusions and Outlook}

Particle and nuclear physics experiments produce data at such volumes that analysis is impossible without some kind of sacrifice. The traditional approach is to keep full records of a small subset of the data, while newer strategies store partial information about a larger fraction. In some cases, however, neither strategy is sufficient to discover and characterize the physics behind evidence of a new kind of particle or interaction. We propose an alternative approach to address this limitation by automatically compressing the full experimental information, relying on an optimal-transport-based Variational Autoencoder with a parameter $\beta$ that controls the bandwidth. 
In a benchmark test, we are able to distinguish between new physics scenarios at a level significantly beyond what is available from the partial event system. While this benchmark contains many salient features of a realistic challenge at the LHC, a complete implementation for the di-muon and other applications requires additional study. For any particular model of new physics, a trigger-level analysis can record an acceptable number of relevant features (e.g. properties of hadronic jets, missing energy and/or lepton four-vectors).  Our approach may provide additional or complementary model-agnostic sensitivity.  This is particularly plausible in models featuring dark showers, which are difficult to distinguish from multijet backgrounds using standard observables (see e.g.~\cite{Strassler:2008bv,Albouy:2022cin}). We leave this for further studies.

In summary, trigger-level analysis methods may be essential to discover and diagnose new physics in the remaining LHC data an the $\beta$-parameterized Variational Autoencoder could be an important tool for extending this physics program in new directions.


\section*{Acknowledgements}
We thank Jesse Thaler for suggesting the parameterized loss function as an alternative to annealing.  DW/EH, BN/SK, JC were supported by the U.S. Department of Energy, Office of Science under grant DE-SC0009920 and contracts DE-AC02-05CH11231 and DE-AC02-76SF00515, respectively. Part of this work was performed at the Aspen Center for Physics, which is supported by National Science Foundation grant PHY-1607611. This research made use of the Matplotlib~\cite{Hunter:2007}, Jupyter~\cite{Kluyver2016jupyter}, NumPy~\cite{harris2020array}, and SciPy~\cite{2020SciPy-NMeth} software packages.


\printbibliography

\section*{Appendix}
\subsection*{Simulation details\label{app:simulation}}
Our first signal benchmark is that of single, real scalar field ($S$) which mixes with the SM Higgs boson. As as result of this mixing, there exist a penguin diagram which induces an exotic decay mode of $B$-mesons to $S$ plus a strangeness-carrying final state \cite{Willey:1982ti,CHIVUKULA198886,Grinstein:1988yu}. Despite the loop-suppression of this partial width, the branching ratio may nevertheless be appreciable due to the very small total width of the $B$-meson. The branching ratio for $S\to 2\mu$ is rather uncertain, but is thought to be between 10\% and 1\% for most of the mass range where $2m_\mu<m_S<m_B-m_K$ \cite{Winkler:2018qyg}. Phenomenologically, the main signature of the model is a low mass (displaced) dimuon resonance, which often sits inside a b-jet. For the purpose of our study, we set $m_S=0.4$ GeV and work with a leading order inclusive unweighted $b$-$\bar b$ sample subject to $p_{\textrm T}>10$ GeV at the parton-level.

Our second signal benchmark is a hidden valley model \cite{Strassler:2006im}, in which the SM Higgs decays to a dark sector with some confining dynamics. This includes a parton shower in the dark sector, followed by a hadronization into dark mesons. Some of these dark mesons are taken to be stable and therefore invisible, while others can decay to a pair of dark photons through a dark sector chiral anomaly. These dark photons may subsequently decay to SM final states, among which dimuons. (See \cite{Knapen:2021eip} for a detailed description and definition of the model.) We take the dark photon mass to be 0.4 GeV, to match the dimuon invariant mass chosen for our first benchmark. The signature of this model is a low mass dimuon resonance, surrounded by hadronic activity plus a small amount of missing energy. The kinematics of the muon pair is very similar in both models, but the pattern of hadronic activity and missing energy in the event can differ substantially. 

While analysis in ~\cite{CMS:2019buh} is able to record hadronic activity in a coarse-grained manner, by recording anti-$k_{\textrm T}$ jets~\cite{Cacciari:2008gp,}, our approach is able to retain more fine-grained substructure information, which may ultimately be needed to distinguish similarly looking signatures.

\subsection*{VAE architecture and training details\label{app:VAEdeets}}

In our experiments, the encoder network consists of four 1D convolution layers with filter size 1024, kernel size 1, and stride 1, followed by a sum layer, followed by four dense layers of size 1024. $\beta$ is also provided as an input to these dense layers, in parallel with the sum layer. Unless otherwise specified, all layers have activation function Leaky ReLU with negative slope coefficient of 0.1. 256 latent space $\mu, \log \sigma^2$ are encoded with linear activation. Sampled codes $z$ are used as input for a decoder, alongside $\beta$. The decoder consists of five layers with size 1024, followed by a linear dense layer which outputs fifty particles represented as $\{(p_{\textrm T}/H_{\textrm T}, \eta, \sin \phi, \cos \phi)\}$, and then an $\arctan$ function reduces this to $\{(p_{\textrm T}/H_{\textrm T}, \eta, \phi)\}$. The explicit $\arctan$ allows the network to avoid learning a discontinuity in $\phi$, which is also the motivation for the trigonometric form of the inputs.

In order to reduce the number of particles to less than 50 for computational speed, events that have more than 50 particles are reclustered using the exclusive-$k_T$ algorithm~\cite{Catani:1993hr} with radius parameter $R = 1.0$ and terminating when there are exactly 50 particles left in the event. The VAE works with $p_{\textrm T}$-normalized inputs, and so the $p_{\textrm T}$ for each event is rescaled by $H_{\textrm T} = \sum_i p_{T,i}$ before being input into the encoder. This $H_{\textrm T}$ is recorded separately for each event, and can be used to rescale the event reconstructions.

We trained the VAE on the $b\bar{b}$ background events. Each batch of training data consists of an array of shape $(n_\text{batch}, 50, 4)$, with $n_{\text{batch}} = 100$, which is input to the VAE encoder, and an array of shape $(n_{\text{batch}})$ samples for $\beta$, which is an auxiliary input to the encoder and decoder. These samples are drawn from a log-uniform distribution in the range $\log_{10} \beta \in [-6,1]$. The loss for each training example is also calculated using this $\beta$. Training is performed using the Adam optimizer \cite{kingma2014adam} with learning rate initialized at $3\times10^{-5}$. Each epoch consists of 1000 batches (as opposed to the entire training dataset), and the learning rate is decreased in factors of $\sqrt{10}$ each time the validation loss has not improved in 10 epochs. Training is stopped after validation loss has not improved in 20 epochs. This entire cycle is repeated 10 times.

The code for this paper can be found at \url{https://github.com/erichuangyf/EMD_VAE/tree/neurips_ml4ps}.

\subsection*{Classifier architecture and training details}

The PFN model used to classify reconstructed signals consists of three time-distributed dense layers of size 128 followed by a sum layer followed by three dense layers of size 128. Each PFN was trained on events  with shape $(n_\text{batch}, 50, 3)$, with the batch size $n_\text{batch} = 1000$. Training is performed using the Adam optimizer with learning rate initialized at $10^{-3}$. The learning rate is decreased in factors of $10^{1/4}$ each time the validation loss has not improved in 5 epochs. Training is stopped after 200 epochs or the validation loss has not improved in 10 epochs.

We also trained the PFN classifier on the hardest two muons taken from the two signal event samples. The training procedure is similar to the PFN on full signal events, except the input shape for each event was truncated to $(n_\text{batch}, 2, 3)$, with the batch size $n_\text{batch} = 1000$. The muon-only PFN model consists of three time-distributed dense layers of size 128 followed by a sum layer followed by three dense layers of size 128. Training is performed using the Adam optimizer with learning rate initialized at $10^{-3}$. The learning rate is decreased in factors of $10^{1/4}$ each time the validation loss has not improved in 5 epochs. Training is stopped after 200 epochs or the validation loss has not improved in 10 epochs.

All training was performed on NVidia Quadro RTX 6000 with CUDA version 11.2 using TensorFlow version 2.4.1. The training time for the VAE model was about 24 hours and the training time for each PFN model was about 5 minutes based on this setup.

\end{document}